\documentclass[prd,aps,floats,twocolumn,showpacs]{revtex4}
\usepackage[dvips]{epsfig}
\begin{document}
\voffset 15mm
\title{Glauber dynamics of phase transitions: 
       SU(3) lattice gauge theory.}

\author{ Alexei Bazavov$^{\rm \,a,b}$, Bernd A. Berg$^{\rm \,a,b}$
and Alexander Velytsky$^{\rm \,c}$}

\affiliation{ $^{\rm \,a)}$ Department of Physics, Florida State 
University, Tallahassee, FL 32306-4350, USA\\
$^{\rm \,b)}$ School of Computational Science, 
Florida State University, Tallahassee, FL 32306-4120, USA\\
$^{\rm \,c)}$ Department of Physics and Astronomy, UCLA, Los
Angeles, CA 90095-1547, USA } 

\date{Apr 29, 2006}

\begin{abstract}
Motivated by questions about the QCD deconfining phase transition,
we studied in two previous papers Model A (Glauber) dynamics of 2D 
and 3D Potts models, focusing on structure factor evolution under 
heating (heating in the gauge theory notation, i.e., cooling of the 
spin systems). In the present paper we set for 3D Potts models (Ising 
and 3-state) the scale of the dynamical effects by comparing to 
equilibrium results at first and second order phase transition
temperatures, obtained by re-weighting from a multicanonical ensemble. 
Our finding is that the dynamics entirely overwhelms the critical and 
non-critical equilibrium effects.

In the second half of the paper we extend our results by investigating 
the Glauber dynamics of pure SU(3) lattice gauge on $N_{\tau}\,
N_{\sigma}^3$ lattices directly under heating quenches from the 
confined into the deconfined regime. The exponential growth factors 
of the initial response are calculated, which give Debye screening mass 
estimates. The quench leads to competing vacuum domains of distinct 
$Z_3$ triality, which delay equilibration of pure gauge theory forever, 
while their role in full QCD remains a subtle question. As in spin
systems we find for pure SU(3) gauge theory a dynamical growth of
structure factors, reaching maxima which scale approximately with the 
volume of the system, before settling down to equilibrium. Their
influence on various observables is studied and different lattice
sizes are simulated to illustrate an approach to a finite volume 
continuum limit. Strong correlations are found during the dynamical
process, but not in the deconfined phase at equilibrium.

\end{abstract}
\maketitle



\section{Introduction}

In investigations of the QCD deconfining phase transition (or 
crossover) by means of heavy ion experiments, one ought to be 
concerned about non-equilibrium effects due to the {\it rapid heating} 
of the system.  With this in mind we have investigated in previous 
papers \cite{BHMV,BMV} the Model~A \cite{ChLu97} (Glauber) dynamics 
of 2D and 3D Potts models. Model~A dynamics includes all diffusive
stochastic local updating schemes (Metropolis, heatbath, etc.) and not 
only the process introduced in \cite{Gl63}. In 3D Potts models spins 
provide degrees of freedom, which mimic Polyakov loops effectively 
\cite{SY82}, while in 2D analytical results \cite{Ba73} allow to check 
on the accuracy of the employed numerical methods. For other approaches
to simulate non-equilibrium quantum fields see Ref.~\cite{BS05}. 

The QCD high temperature vacuum is characterized by ordered Polyakov
loops, which are similar to spins in the low temperature phase of
the 3D 3-state Potts model. We model heating by a quench from the
disordered into the ordered phase, which thus corresponds to a
cooling quench in the analogue spin model. Time evolution after
the quench leads to vacuum domains of distinct triality under the
Z$_3$ center of the SU(3) gauge group. It appears that these competing 
domains are the underlying cause for the explosive growth of structure 
factors $F_i(t)$, which we encounter in the time evolution after a 
heating quench. We use the term {\it spinodal decomposition} loosely
to denote generically such a time period of globally unstable behavior.

Relaxation of the system at its new temperature becomes only feasible 
after each structure factor has overcome its maximum value. While the 
maximum value of the structure factor diverges with lattice size, its 
initial and final equilibrium values are finite in the normalization 
chosen in the paper. The time (measured in updates per degree of 
freedom) for reaching the maximum diverges with lattice size unless
the underlying order-order symmetry is broken. 
Once the system has equilibrated at high temperature, the subsequent
temperature fall-off is driven by spatial lattice expansion and the
system stays in quasi-equilibrium during this period. So one has
different time scales under heating and cooling~\cite{TV05}.

The early time evolution of SU(3) gauge theory after the quench is 
well described by stochastic equations, which follow from dynamical 
generalizations of equilibrium Landau-Ginzburg effective action models. 
We calculate the exponential growth factor of this linear approximation 
and use a phenomenological model~\cite{MiOg02} to estimate the Debye 
screening mass for two temperatures above the deconfining $T_c$. 

Finally we compare measurements of Polyakov loop correlations, gluonic
energy densities and pressures around structure function maxima with
their equilibrated values in the deconfined region at high temperatures.
These measurements are of interest for a scenario in which the heating
process turns back to cooling before actually reaching the equilibrium
side of the structure factor maxima. In the conclusions we continue
this discussion.

In the next section we introduce our notations and some preliminaries.
Section~\ref{sec_Potts} deals with Potts models. First equilibrium
properties of structure functions are established by means of
multicanonical simulations of the 3D Ising and 3-state Potts model.
Subsequently their dynamical evolution after a quench is investigated,
extending previous results. In section~\ref{sec_SU3} we present our
simulations of pure SU(3) lattice gauge theory.  Some SU(3) data were 
already reported at the 2004 APS DPF conference~\cite{BBV05}. As these 
simulations are very CPU time consuming it took over one more year to 
collect the present statistics. Summary and conclusions are given in
the final section~\ref{sec_sum}.

\section{Notation and Preliminaries \label{sec_notation}}

We summarize our basic notations and concepts in this section.

\subsection{Models \label{sec_models}}

We simulate $q$-state Potts models with the energy function
\begin{equation} \label{E_Potts}
  E\ =\  2 \sum_{\langle ij\rangle} 
  \left(\frac{1}{q}-\delta_{q_iq_j}\right)
\end{equation}
where the sum is over nearest neighbors of a hypercubic lattice in
D dimensions. The spins $q_i$ of the system take on the values
$q_i=0,\dots,q-1$. The factor of two and the term $1/q$ is introduced 
to match for $q=2$ with Ising model conventions \cite{BBook}. 
Simulations are carried out with the Boltzmann factor $\exp(-\beta E)$.

The Wilson action for pure SU(3) non-Abelian Euclidean lattice gauge 
theory is
\begin{equation}
  S_A=\frac{2\cdot3}{g^2}\sum_{n,\mu\nu}[1-\frac1{2\cdot3}\mbox{Tr}
  (U_{n,\mu\nu}+\mbox{h.c.})],
\end{equation}
where $U_{n,\mu\nu} = U_{n,\mu}U_{n+\hat{\mu},\nu}
U^\dagger_{n+\hat{\nu},\mu} U^\dagger_{n,\nu}$ denotes the product of 
the SU(3) link matrices in the fundamental representation around a 
plaquette and the sum runs over all plaquettes. Simulations are 
done with the Boltzmann factor $\exp(S_A)$.

The Markov chain Monte Carlo (MC) process provides model~A
(Glauber) dynamics in the classification of Ref.~\cite{ChLu97}.
For Potts models we use the heatbath algorithm of \cite{BBook}
and for SU(3) gauge theory the Cabibbo-Marinari \cite{CaMa82} 
heatbath algorithm and its improvements of Ref.~\cite{FaHa84} 
(no over-relaxation, to stay in the universality class of Glauber 
dynamics). In each case a time step is a sweep of systematic updating 
through the lattice, which touches each degree of freedom once. With
small statistics we have checked that updating in random order gives
similar results up to a slowing down of the evolution speed by a 
constant factor. This is expected as in equilibrium simulations
random updating has larger autocorrelations than systematic 
updating~\cite{BBook}. For our equilibrium simulations of Potts 
models we used a multicanonical \cite{BN92} Metropolis algorithm.

\subsection{Structure Factors \label{sec_sf}}

Consider two-point correlation functions defined by
\begin{equation}\label{e:L0LR_def}
    \langle u_0(0)u_0^\dagger(\vec{j})\rangle_L=\frac{1}{N_\sigma^3}
    \sum_{\vec{i}}u_0(\vec{i})u_0^\dagger(\vec{i}+\vec{j}),
\end{equation}
where 
$\vec{i}$ denotes spatial coordinates. Periodic boundary conditions
are used and the subscript $L$ on the left-hand side reminds us that
the average is taken over the spatial lattice.
For gauge systems we deal with fluctuations of the Polyakov loop,
for analogue spin systems with fluctuations of the magnetization.

The finite volume continuum limit of (\ref{e:L0LR_def}) is achieved by
lattice spacing $a\rightarrow0$ and $N_\sigma\rightarrow\infty$ with 
the physical length of the box $L=aN_\sigma=\rm const$. This means that
\begin{equation}\label{e:L0LR_contlim}
    \langle u_0(0)u_0^\dagger(\vec{j})\rangle_L=\frac{1}{a^3N_\sigma^3}
    \sum_{\vec{i}}a^3 u_0(\vec{i})u_0^\dagger(\vec{i}+\vec{j})
\end{equation}
transforms into
\begin{equation}\label{e:LOLR_contlim2}
    \langle u(0)u^\dagger(\vec{R})\rangle_L=\frac{1}{L^3}
    \int d^3r\,u(\vec{r})u^\dagger(\vec{r}+\vec{R}),
\end{equation}
with $\vec{r}=a\vec{i}$, $\vec{R}=a\vec{j}$, $u(\vec{r})=u_0(\vec{i})$, 
and so on. We define the structure function $F(\vec{p})$ as Fourier 
transformation of the two-point correlation function 
(\ref{e:LOLR_contlim2}):
\begin{equation}\label{e:SF2_def}
    F(\vec{p})=\int \langle u(0)u^\dagger(\vec{R})\rangle_L\,
    e^{i\,\vec{p}\,\vec{R}}\, d^3R.
\end{equation}
Periodic boundary conditions imply:
\begin{equation}\label{e:k_def}
   \vec{p} = \frac{\vec{k}}{a} = \frac{2\pi}{L}\vec{n}\,,
\end{equation}
where $\vec{n}$ is an integer vector $(0,0,0)$, $(0,0,1)$, and so on.
The discretized version of (\ref{e:SF2_def}) is
\begin{equation}\label{e:SF2_disc}
    F(\vec{p})=\sum_{\vec{j}}a^3\,\langle u_0(0)u_0^\dagger(\vec{j})
    \rangle_L\, e^{i\,\vec{k}\,\vec{j}}.
\end{equation}
Using the definition (\ref{e:L0LR_def}) and shifting the $\vec{j}$
summation one arrives (after straightforward algebra) at the expression
\begin{equation}\label{SFsquare}
   F(\vec{p})= \frac{a^3}{N_\sigma^3}\left|\, \sum_{\vec{i}}
   e^{-i\,\vec{k}\,\vec{i}}\, u_0(\,\vec{i}\,) \,\right|^2\,,
\end{equation}
where we may rewrite the product in the exponent as
\begin{equation}\label{e:aki}
    a\,\vec{p}\,\vec{i}\,=\,\vec{k}\,\vec{i}\,=
    \frac{2\pi}{N_\sigma}\,\vec{n}\,\vec{i}\ .
\end{equation}
As we let the system evolve after a quench $u_0(\vec{i})$ becomes 
time-dependent: $u_0(\vec{i},t)$. The time $t$ corresponds to the
dynamical process, i.e., in our case the Markov chain model~A dynamics.
We consider an ensemble of systems (replica) and dynamical observables 
are calculated as ensemble averages denoted by $\langle...\rangle$.
The time-dependent structure functions averaged over replicas are:
\begin{equation}\label{ens_aver_S}
  F_{\vec{p}}(t)\,=\,\left\langle F(\vec{p},t)\right\rangle\ .
\end{equation}
During our simulations they are averaged over rotationally equivalent 
momenta and the notation 
\begin{equation} \label{sf}
  F_i(t)
\end{equation}
is used for the structure function at momentum
\begin{equation} \label{momenta}
  \vec{p} = \frac{\vec{k}}{a} = {2\pi\over L}\,\vec{n}
\end{equation}
where $|\vec{n}|=n_i$ defines $i$.
The $F_i$ are called structure function modes or structure factors 
(SFs). We recorded the following modes (including the permutations) 
$n_1$: $(1,0,0)$, $n_2$: $(1,1,0)$, $n_3$: $(1,1,1)$, $n_4$: $(2,0,0)$, 
$n_5$: $(2,1,0)$, $n_6$: $(2,1,1)$, $n_7$: $(2,2,0)$, $n_8$: $(2,2,1)$ 
and $(3,0,0)$, $n_9$: $(3,1,0)$. Note that there is an accidental 
degeneracy in length for $n_8$. We measured also higher modes, in
some cases up to $n_{64}$. They exhibit the same behavior as the
lower modes, but the data are far more noisy, so that we abstain
from reporting these results.
A difference to the normalization of \cite{BHMV,BMV} is that in the
present paper we average over the permuted momenta instead of just
summing them up. For instance, for the $F_1$ SF the difference is
a multiplicative factor of three.

\section{Potts Models \label{sec_Potts}}

For the analogue spin models the lattice spacing $a$ cannot be varied.
We set $a=1$, so that the distinction between $L$ and $N_{\sigma}$,
$\vec{p}$ and $\vec{k}$ becomes superfluous. We use $L$ and $\vec{k}$ 
in the following. The normalization of the SFs differs from our previous 
work \cite{BHMV,BMV}. It is chosen so that they approach constant values 
in the infinite volume limit of equilibrium simulations of spin systems 
at {\it non-}critical temperatures. This follows from the fact that the
random fluctuations in (\ref{sf}) are of order $\sqrt{V}=\sqrt{L^3}$.
At a critical temperature of a second order phase transition a 
divergence of the SFs is then encountered
as we illustrate for the 3D Ising model. A sustained increase of a 
SF with lattice size cannot be stronger than being proportional to 
$V=L^3$, because an upper bound on each SF is obtained by setting all 
values in the sum of Eq.~(\ref{SFsquare}) equal to one. 

\subsection{Equilibrium Results \label{sec_equil}}

In this section we compile SF estimates from equilibrium simulations 
of the 3D Ising and 3-state Potts model on $L^3$ lattices. Our 
simulations are carried out in a multicanonical ensemble \cite{BN92}, 
covering a temperature range from $\beta_{\min}=0$ (infinite 
temperature) to $\beta_{\max} > 0$ below the phase transition 
temperature of the respective system. Instead of relying on a 
recursion (see, e.g., \cite{BBook}), the multicanonical parameters
were extracted by finite size (FS) extrapolation from smaller to 
larger system, which is an efficient way when the FS behavior is
controllable. 

The advantage of using multicanonical simulations is that accurate 
values of the SF peaks can be determined from one data set. 
Re-weighting of a canonical simulation \cite{FeSw88} allows accurate 
determination of the maxima of one quantity, but on finite lattices 
the maxima of different observables are too far apart to be within the
re-weighting range of one canonical simulation. We find it convenient 
to have the entire range of interest covered in one simulation. In 
particular equilibration of the configurations 
around the transition and in the ordered phase is then secured due to 
frequent excursions into the discorded region all the way to $\beta=0$.

\subsubsection{3D Ising Model \label{sec_3DI}}

At the critical point the two-point function on an infinite lattice 
falls off with a power law, which defines the critical exponent $\eta$:
\begin{equation} \label{2ptfct}
  f(\vec{x}) \,=\, \langle\,s(\vec{0})\,s(\vec{x})\,\rangle \,\sim\,
  |\vec{x}|^{-d+2-\eta}\,,~~~|\vec{x}|\to \infty\, .
\end{equation}
This determines the low-momentum behavior of the Fourier 
transformation $F(\vec{k})$. Namely,
\begin{eqnarray}  \nonumber
  \hat{F}(\lambda\vec{k}) &\sim& \int d^dx\,
  e^{i\,\lambda\,\vec{k}\cdot\vec{x}}\,|\vec{x}|^{-d+2-\eta} \\
  &=& \int \frac{d^dx'}{\lambda^d}\, e^{i\,\vec{k}\cdot\vec{x'}}\,
  \lambda^{d-2+\eta}\,|\vec{x'}|^{-d+2-\eta} \nonumber \\
  &\sim& \lambda^{\eta-2}\,\hat{F}(\vec{k}) \nonumber
\end{eqnarray}
holds and, therefore,
\begin{equation} \label{sfcritical}
  \hat{F}(\vec{k}) \sim |\vec{k}|^{-b}\,,~~b=2-\eta\,,
  ~~{\rm for}~~|\vec{k}|\to 0\,.
\end{equation}
For fixed $\vec{n}$ we have $\vec{k}=2\pi\vec{n}/L$ and we find
for any fixed value of $\vec{n}$ the finite size scaling (FSS) 
divergence 
\begin{equation} \label{sf_fs}
  \hat{F}_{\vec{n}} \,\sim\, L^b\,,~~b={2-\eta}\,,
  ~~{\rm for}~~L\to\infty
\end{equation}
with lattice size.

\begin{table}[tf]
\caption{Statistics and SF maxima $F_1^{\max}$ at $\beta_m$ from our 
equilibrium simulations of the 3D Ising model on $L^3$ lattices.
\label{tab_3DI}}
\medskip
\centering
\begin{tabular}{|c|c|c|c|c|c|}   \hline
$L$& sweeps                 & $F_1^{\max}$    & $\beta_m$& cycles
\\ \hline
20 & $32\times 5\cdot 10^4$ & $\ 17.00\,(26)$ & 0.219874 & 83 \\ \hline
30 & $32\times 2\cdot 10^5$ & $\ 37.97\,(70)$ & 0.220825 & 89 \\ \hline
44 & $32\times 6\cdot 10^5$ & $\ 78.4\,(1.6)$ & 0.221146 & 70 \\ \hline
56 & $32\times 1\cdot 10^6$ & $ 129.6\,(2.7)$ & 0.221345 & 45 \\ \hline
66 &$32\times 1.6\cdot 10^6$& $ 175.6\,(4.9)$ & 0.221387 & 43 \\ \hline
80 &$64\times 2\cdot 10^6\times 3$&$257.4\,(2.4)$&0.221462 &$65+72+67$
\\ \hline
\end{tabular} \end{table} 

The infinite volume phase transition temperature of the 3D Ising model 
is estimated to be $\beta_c = 0.22157\,(3)$, e.g., see \cite{3DI_Tc}.
In our multicanonical simulations we cover the range from 
$\beta_{\min}=0$ to $\beta_{\max}=0.25$, well including the transition 
region. Table~\ref{tab_3DI} gives an overview of the lattice sizes
and the accumulated statistics (a sweep updates each spin once) 
together with our estimate of the maximum values $F_1^{\max}$ of the 
first SF, evaluated at the value $\beta_m$.  Error bars are given in 
parenthesis and apply to the last digits of the number in front. They 
are calculated with respect to a number of jackknife bins given by 
the first number in column two of the table, and the multicanonical 
re-weighting procedure uses the logarithmic coding described 
in~\cite{MUCArwght}. Three independent runs were carried out for 
the $L=80$ lattice. Before starting with measurements we normally
performed the number of sweeps of one measurement bin for
reaching equilibrium. This is sufficient because equilibration
problems are mild in multicanonical simulations. Running time 
for each of our $L=80$ simulations was about three months on a 
2~GHz Athlon PC. The last column of table~\ref{tab_3DI} gives the 
number of cycles
$$ (\beta_e\le\beta_{\min}) \to (\beta_e\ge\beta_{\max}) \to  
   (\beta_e\le\beta_{\min})\,, $$
which the Markov process performed during the production run,
where $\beta_e$ is the effective energy-dependent $\beta$ of the
multicanonical procedure. 

\begin{figure}[-t] \begin{center} 
\epsfig{figure=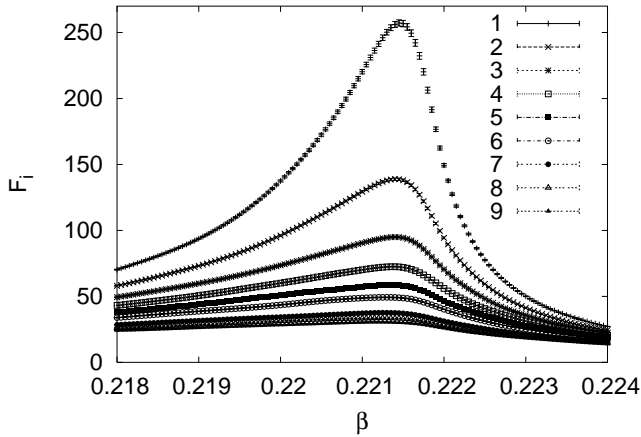,width=\columnwidth} \vspace{-1mm}
\caption{SFs $F_i,\, i=1,\dots,9$ from simulations of the Ising model 
on an $80^3$ lattice. \label{fig_3DI80sf} }
\end{center} \vspace{-3mm} \end{figure}

\begin{figure}[-t] \begin{center} 
\epsfig{figure=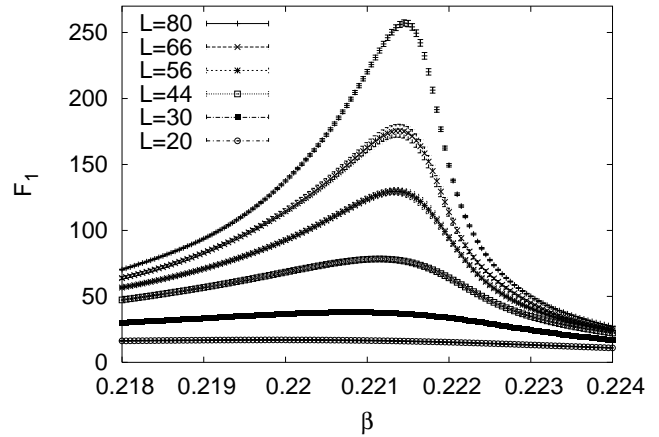,width=\columnwidth} \vspace{-1mm}
\caption{Finite size behavior of SF $F_1$ from Ising model simulations
on $L^3$ lattices. \label{fig_3DIsf1} }
\end{center} \vspace{-3mm} \end{figure}

\begin{figure}[-t] \begin{center} 
\epsfig{figure=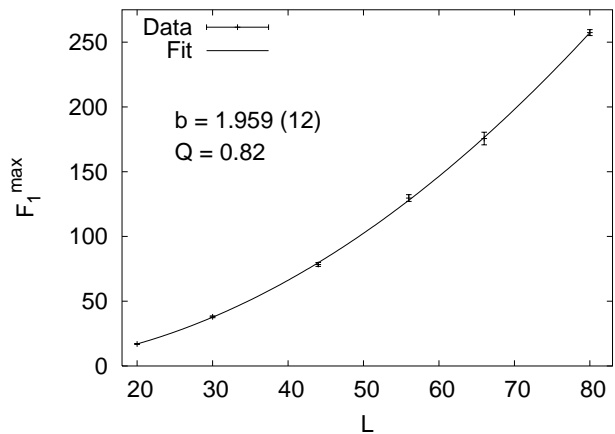,width=\columnwidth} \vspace{-1mm}
\caption{Fit of the $F_1^{\max}$ maxima of table~\ref{tab_3DI} to 
the FSS form~(\ref{sf_fs}). \label{fig_3DIsf1fit} }
\end{center} \vspace{-3mm} \end{figure}

For our largest lattice the SFs 1-9 are plotted in 
Fig.~\ref{fig_3DI80sf}, where we restrict $\beta$ to a neighborhood
of the critical temperature. Each SF develops a clear peak, only that 
the peaks for the higher SFs are less pronounced than those for the 
lower. In particular the scale of the figure does not resolve the 
peaks for the SFs $\ge 7$ anymore. These peaks are found on a reduced
scale and for each SF the FSS behavior (\ref{sf_fs}) holds. However, 
the numerical accuracy decreases with increasing $\left|\vec{k}\right|$. 
So we are content with simply analyzing the FSS behavior of SF~1. 
Fig.~\ref{fig_3DIsf1} shows SF~1 for all our lattice sizes and the 
maxima values are collected in table~\ref{tab_3DI}. A two parameter fit 
to the form~(\ref{sf_fs}) is shown in Fig.~\ref{fig_3DIsf1fit}. It 
gives $b=1.959\,(12)$ with a goodness of fit $Q=0.82$ (for the 
definition of $Q$ see, e.g., Ref.~\cite{BBook}), a result well 
compatible with the high precision estimates $\eta=0.0364\,(5)$ 
given in the review article~\cite{PeVi02} on critical phenomena and 
renormalization group theory.

\subsubsection{3D 3-state Potts Model \label{sec_3DP3q}}

\begin{table}[tf]
\caption{Statistics and SF maxima $F_1^{\max}$ at $\beta_m$ for our 
equilibrium simulations of the 3D 3-state Potts model on $L^3$ 
lattices.  \label{tab_3DP3q}} \medskip
\centering
\begin{tabular}{|c|c|c|c|c|c|}   \hline
$L$& sweeps                  & $F_1^{\max}$ &$\beta_m$& cycles
\\ \hline
20 &$\ 32\times 1.2\cdot 10^5$&$19.00\,(22)$& 0.274273 &\ 59\\ \hline
30 &$\ 32\times 5.2\cdot 10^5$&$38.11\,(41)$& 0.274924 &\ 71\\ \hline
40 &$\ 32\times 1.5\cdot 10^6$&$60.30\,(50)$& 0.275116 &\ 73\\ \hline
50 & $126\times 1.5\cdot 10^6$&$80.46\,(55)$& 0.275181 & 131\\ \hline
\end{tabular} \end{table} 

For the 3D 3-state Potts model one deals with a relatively weak
first order phase transition at $\beta_c = 0.2752720\,(49)$, a 
value which averages two somewhat inconsistent ($Q=0.003$ for the
Gaussian difference test) estimates of the literature~\cite{KaSt00} 
(because of the inconsistency the error bars are averaged here and
not reduced). In our multicanonical simulations we cover the 
range from $\beta_{\min}=0$ to $\beta_{\max}=0.33$. An overview of
the statistics and some results are given in table~\ref{tab_3DP3q},
similarly as before for the 3D Ising model in table~\ref{tab_3DI}.

\begin{figure}[-t] \begin{center} 
\epsfig{figure=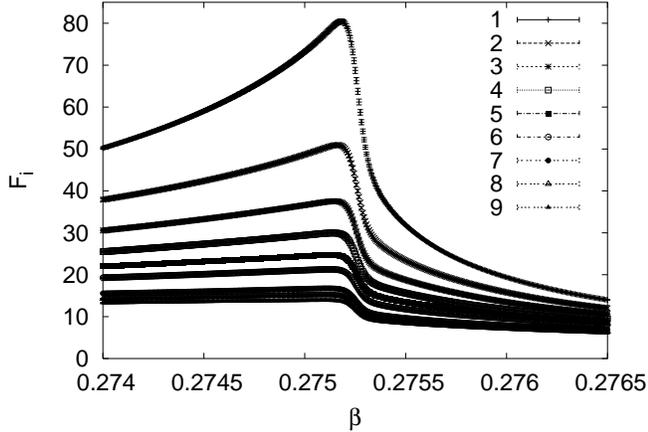,width=\columnwidth} \vspace{-1mm}
\caption{SFs $F_i,\ i=1,\dots,9$ from simulations of the 3-state Potts 
model on a $50^3$ lattice. \label{fig_3DP3q50sf} }
\end{center} \vspace{-3mm} \end{figure}

\begin{figure}[-t] \begin{center} 
\epsfig{figure=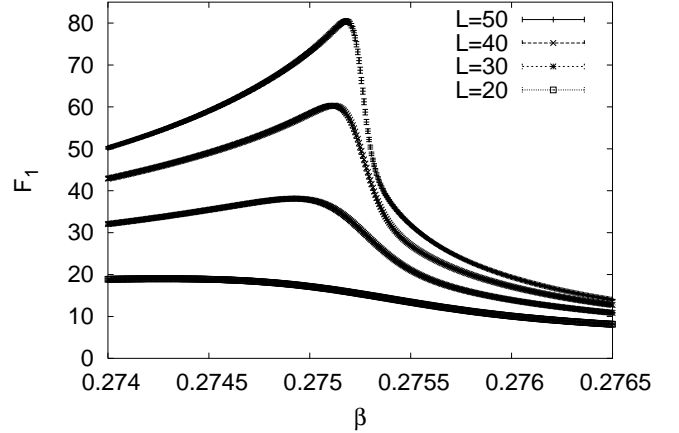,width=\columnwidth} \vspace{-1mm}
\caption{Finite size behavior of SF $F_1$ from 3-state Potts model 
simulations on $L^3$ lattices. \label{fig_3DP3qsf1} }
\end{center} \vspace{-3mm} \end{figure}

For our $50^3$ lattice the SFs 1-9 are plotted in 
Fig.~\ref{fig_3DP3q50sf}, where we restrict $\beta$ to a neighborhood
of the transition temperature. As for the 3D Ising model each SF 
develops a clear peak, but the shapes are significantly different.
A relatively smooth increase is followed by a rather abrupt decrease.
The lattice size dependence of SF~1 is depicted in 
Fig.~\ref{fig_3DP3qsf1}, which indicates (as expected) that 
the abrupt decrease develops into a discontinuity for $L\to\infty$.
The increase of the structure function maxima is irregular and
smaller from $L=40$ to $L=50$ than from $L=30$ to $L=40$. 
Asymptotically for $L\to\infty$ a finite maximum value is expected 
in case of a first order phase transition. Within our limited lattice 
sizes this is not yet seen, but a power law fit (\ref{sf_fs}) of the 
type of Fig.~\ref{fig_3DIsf1fit}, which is the large $L$ behavior in 
case of a second order transition, becomes entirely inconsistent: 
$Q=2.7\cdot 10^{-11}$ is the goodness of fit obtained.

\subsection{Quenches \label{sec_Quenches}}

\begin{table}[tf]
\caption{Repetitions of quenches from $\beta=0.2$ to $\beta_f$
for the 3D 3-state Potts model on $L^3$ lattices.  
\label{tab_3DP3qQuench}} \medskip
\centering
\begin{tabular}{|c|c|c|c|c|c|c|c|}   \hline
$\beta_f\ \backslash\ L:$
            & 20& 40& 60& 80& 100& 120\\ \hline
$0.3~\qquad$&640&640&640&320& 320& 320\\ \hline
$0.27\qquad$&640&320&320& 32&    &    \\ \hline
\end{tabular} \end{table} 

After outlining the equilibrium scenario, let us discuss the time 
evolution after a quench from the disordered into the ordered
phase of the 3D 3-state Potts model. An overview of our statistics 
is given in table~\ref{tab_3DP3qQuench}. We quench from $\beta =0.2$ 
to the $\beta_f$ value given in the table, which collects the 
numbers of repetitions of each quench. Error bars are then 
calculated with respect to 32 bins. Larger lattices exhibit 
self-averaging, so that one needs less repetitions than for smaller 
lattices.

\begin{figure}[-t] \begin{center} 
\epsfig{figure=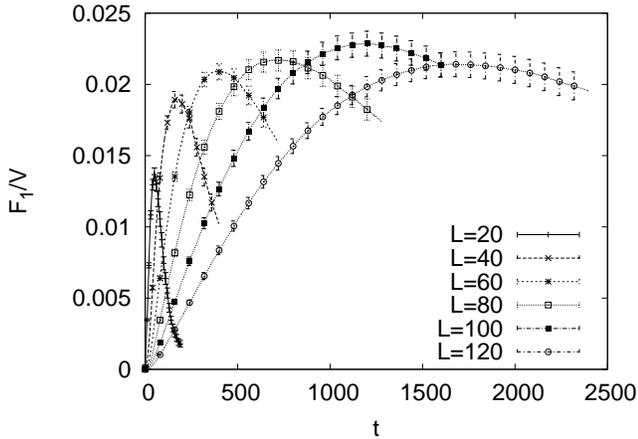,width=\columnwidth} \vspace{-1mm}
\caption{Time evolution of SF $F_1$ for the 3-state Potts model on 
$L^3$ lattices after a quench from $\beta =0.2\to 0.3$. 
\label{fig_3DP3qsf1b30} }
\end{center} \vspace{-3mm} \end{figure}

In previous work~\cite{BMV} we have investigated the quench $\beta = 
0.2 \to 0.3$ and its subsequent stochastic time evolution on lattices
up to size $80^3$. Meanwhile we have extended the SF part of this 
investigation to lattices of size up to $120^3$ and 
Fig.~\ref{fig_3DP3qsf1b30} shows the time evolution of SF~1 after 
this quench. Note that we divide the SF by an extra volume factor 
in this figure. So its initial increase with lattice size is faster 
than $\sim V$, the maximum sustained increase we discussed in the
first paragraph of section~\ref{sec_Potts}. For our largest lattices, 
$L\ge 80$, the increase appears to level off to precisely 
\begin{equation} \label{sfL_asymptotic}
  F_1^{\max}(L)\ \sim\ V = L^3~~~{\rm for}~~~L\to\infty
\end{equation}
Heuristically this behavior during spinodal decomposition may be 
expected: The quench changes the temperature in the entire lattice 
instantaneously. It is then plausible that the local contribution 
to the SF is, in the average, everywhere the same. So one expects an 
increase $\sim V$ of the maxima. The initial overshooting may be 
explained by an increase of correlations with lattice size, which 
levels off once the lattice size exceeds the correlation length.

\begin{figure}[-t] \begin{center} 
\epsfig{figure=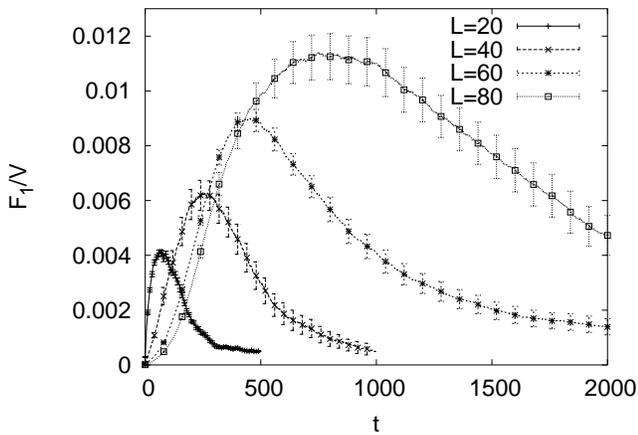,width=\columnwidth} \vspace{-1mm}
\caption{Time evolution of SF $F_1$ for the 3-state Potts model on 
$L^3$ lattices after a quench from $\beta =0.2\to 0.28$. 
\label{fig_3DP3qsf1b28} }
\end{center} \vspace{-3mm} \end{figure}

To test how this growth of the signal proportional to the volume
depends on the depth of the quench into the ordered region, we
performed a quench to a temperature closer to the transition 
temperature, $\beta = 0.2 \to 0.28$. As shown in 
Fig.~\ref{fig_3DP3qsf1b28} we find the same phenomenon
as before: The maximum sustained increase $\sim V$ is initially
overshot. The growth of the signal is weaker than before, 
as is expected since the system does not change so drastically. 

\begin{figure}[-t] \begin{center} 
\epsfig{figure=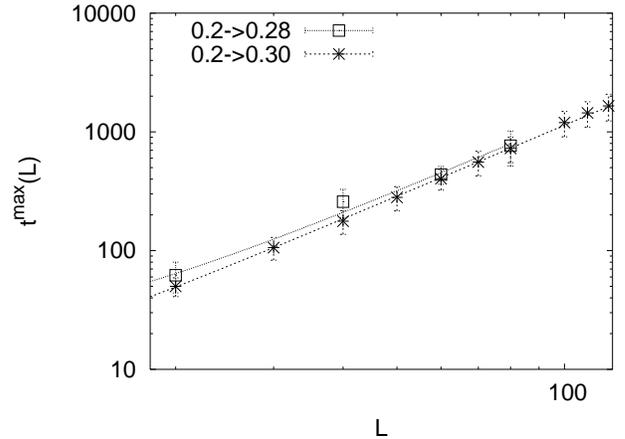,width=\columnwidth} \vspace{-1mm}
\caption{Time positions of the SF $F_1$ maxima versus lattice size.
\label{fig_3DP3qtmax} }
\end{center} \vspace{-3mm} \end{figure}

In both figures we see that the time positions $t^{\max}$ of the 
SF~1 maxima move towards larger values with increasing lattice 
size.  For our two quenches $t^{\max}(L)$ is plotted in 
Fig.~\ref{fig_3DP3qtmax} on a log-log scale. With parameters
$a_0$ and $a_1$ both curves can consistently be fitted to the 
expected form
\begin{equation} \label{tmaxL}
  t^{\max}(L) = a_0 + a_1\,L^2\ .
\end{equation}
As $t$ is measured in units of sweeps, 
the number of spin updates per time unit does not depend on $L$.
In spin systems $t$ is thus proportional to the physical time.
After the quench into the ordered phase the infinite spin system
cannot be equilibrated in any finite time, a fact known in condensed 
matter physics~\cite{equilibration}. The explanation for this 
phenomenon is that the systems grows initially competing domains of
three distinct orientations. To dissolve these domains by local random 
fluctuations until one of them dominates the entire lattice is a slow
process, which requires of order $L^2$ time.

Visualization of these domains faces difficulties, because naive 
geometrical definitions do not work. Compare Fig.~8 of Ref.~\cite{BMV}.
For analogue Potts models this is overcome by the Fortuin-Kasteleyn 
\cite{FK72} cluster definition, but there is no immediate 
generalization to gauge theories, although promising ideas have 
been published \cite{Fo03}. Here we do not investigate this question 
any further. We think that it is safe to assume that competing domains 
are in both, spin and gauge systems, the underlying cause for the 
explosive growth of structure factors $F_i(t)$, which we encounter 
in their time evolution after a heating quench. 

\begin{figure}[-t] \begin{center} 
\epsfig{figure=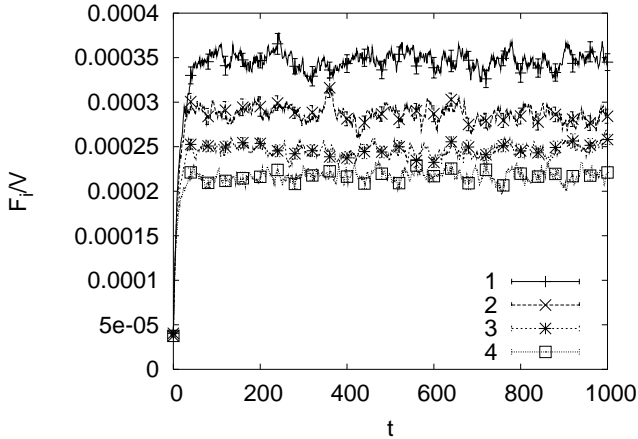,width=\columnwidth} \vspace{-1mm}
\caption{Time evolution of SFs $F_i,\ i=1,\dots,4$ for the 3-state 
Potts model on a $40^3$ lattice after a (non-critical) quench from 
$\beta =0.2\to 0.27$.  \label{fig_3DP3qsf1b27} }
\end{center} \vspace{-3mm} \end{figure}

Finally in this section, based on 640 repetitions 
Fig.~\ref{fig_3DP3qsf1b27} demonstrates that for a non-critical 
quench nothing more than a smooth transition from one equilibrium 
value to the next happens. Therefore the explosive growth of
SFs is a unambiguous signal that $\beta_f$ is indeed in the ordered
phase.

\section{SU(3) \label{sec_SU3}}

We report results from quenches of pure SU(3) lattice gauge theory on 
$N_{\tau}\,N_{\sigma}^3$ lattices. Our statistics is summarized in 
tables~\ref{tab_SU3Quench} and~\ref{tab_SU3kc}. All quenches are from 
the initial value $6/g^2=5.5$. The $4\times N^3_{\sigma}$ simulations 
of table~\ref{tab_SU3Quench} were already reported in~\cite{BBV05}. The 
simulations for the other lattices are new. The difference between the
tables is that for the lattices of table~\ref{tab_SU3Quench} 
we follow the quench all the way to its equilibrium value at $T_f$, 
while for the lattices of table~\ref{tab_SU3kc} we calculated only
the initial increase of the SFs as needed for the determinations
of critical modes in section~\ref{sec_Debye}.

\begin{table}[tf]
\caption{Quenches from $6/g^2=5.5$ to $6/g^2_f$ for pure SU(3) lattice
gauge theory (n denotes the number of repetitions).  
\label{tab_SU3Quench}} \medskip
\centering
\begin{tabular}{|c||c|c|c||c|c|c|c|}   \hline
Lattice    & $T_f/T_c$& $6/g^2_f$ & n & 
             $T_f/T_c$& $6/g^2_f$ & n \\ \hline
$4\times 16^3$& 1.250& 5.802740& 10$\,$000& 1.568& 5.920000& 10$\,$000
\\ \hline
$4\times 32^3$& --   &  --     &  --   & 1.568& 5.920000&\ 4$\,$000
\\ \hline
$4\times 64^3$& --   &  --     &  --   & 1.568& 5.920000&~~\ 170
\\ \hline
$6\times 24^3$& 1.250& 6.022334&\ 6$\,$000& 1.568& 6.165427&\ 6$\,$000
\\ \hline
$8\times 32^3$& 1.250& 6.206036&\ 3$\,$000& 1.568& 6.364572&\ 3$\,$000
\\ \hline
\end{tabular} \end{table} 

\begin{table}[tf]
\caption{Initial quenches from $6/g^2=5.5$ to $6/g^2_f$ for pure SU(3) 
lattice gauge theory (n as in table~\ref{tab_SU3Quench}).  
\label{tab_SU3kc}} \medskip
\centering
\begin{tabular}{|c||c|c|c||c|c|c|c|}   \hline
Lattice    & $T_f/T_c$& $6/g^2_f$ & n & 
             $T_f/T_c$& $6/g^2_f$ & n \\ \hline
$4\times 32^3$& 1.250& 5.802740&\ 3$\,$000&--&  --    &\ -- \\ \hline
$4\times 64^3$& 1.250& 5.802740&\ 140&   -- &   --    &\ -- \\ \hline
$6\times 48^3$& 1.250& 6.022334&\ 600& 1.568& 6.165427&\ 750\\ \hline
$6\times 60^3$& 1.250& 6.022334&\ 200& 1.568& 6.165427&\ 200\\ \hline
$8\times 56^3$& 1.250& 6.206036&\ 400& 1.568& 6.364572&\ 400\\ \hline
\end{tabular} \end{table} 

The new data serve to study the quantum continuum limit $a\to 0$ 
(in physical units like fermi). The final values $g^2_f$ of the bare 
coupling constants are chosen, so that the values of $T_f/T_c$ 
stay at the fixed ratios given in the table. For this we take 
(substantial) corrections to the two-loop equation of Lambda 
lattice into account, which follow from renormalization group
results tabulated in Ref.~\cite{BoEn96}. As the use of
tables is tedious, we like to mention that with an accuracy of 0.5\% 
and better our $T_f/T_c$ values are reproduced by using the formula
\begin{equation} \label{Lambda_L}
  \Lambda_L(g^2)\ =\ \Lambda^{as}_L(g^2)\,\lambda(g^2)
\end{equation} 
where $\Lambda^{as}_L(g^2)$ is given by (e.g.,~\cite{HH80})
$$  \Lambda_L^{as}\ =\ \left(b_0\,g_0^2\right)^{-b_1/(2b_0^2)}\,
    e^{-1/(2b_0\,g^2)} $$
with $b_0=\frac{11}{3}\frac{N_c}{16\pi^2}$, 
$b_1=\frac{34}{3}\left(\frac{N_c}{16\pi^2}\right)^2$ and
$$ \lambda(g^2)\ =\ 1+a_1\,e^{-a_2/g^2}+a_3\,g^2+a_4\,g^4 $$
with $a_1=71553750$, $a_2=19.48099$ $a_3=-0.03772473$, and
$a_4=0.5089052$.

\begin{figure}[-t] \begin{center} 
\epsfig{figure=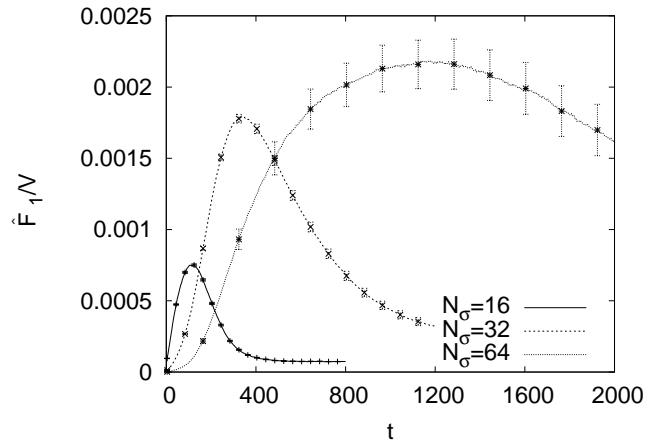,width=\columnwidth} \vspace{-1mm}
\caption{Time evolution of SF $F_1/V$ for SU(3) lattice gauge
theory on $4\times N^3_{\sigma}$ lattices after a quench 
$6/g^2=5.5 \to 5.92$.  \label{fig_SU3sf1} }
\end{center} \vspace{-3mm} \end{figure}

For $N_{\tau}=4$, fixed, Fig.~\ref{fig_SU3sf1} shows the divergence of 
the SF~1 maxima with increasing lattice size $N_{\sigma}^3$ as well as 
a $t^{\max}(N_{\sigma})\sim N_{\sigma}^2$ behavior in complete analogy 
to our results for Potts models. All the lattices of 
Fig.~\ref{fig_SU3sf1}
are quenched to the bare coupling constant $g^2_f=6/5.92$. Therefore
the time scale of the Markov process (determined by the Boltzmann
factors) is the same on all these lattices and up to an unknown
multiplicative factor identified with that of a dissipative, 
non-relativistic dynamics. Non-relativistic does not necessarily 
mean that the propagation of the signal through the lattice is slow.
In the contrary, Galilee transformations set no upper limit on speeds.
Our quench changes the temperature instantaneously through the entire 
lattice, while the subsequent propagation of the response proceeds 
through local interactions.

\subsection{Finite Volume Continuum Limit\label{FVCL}}

\begin{figure}[-t] \begin{center} 
\epsfig{figure=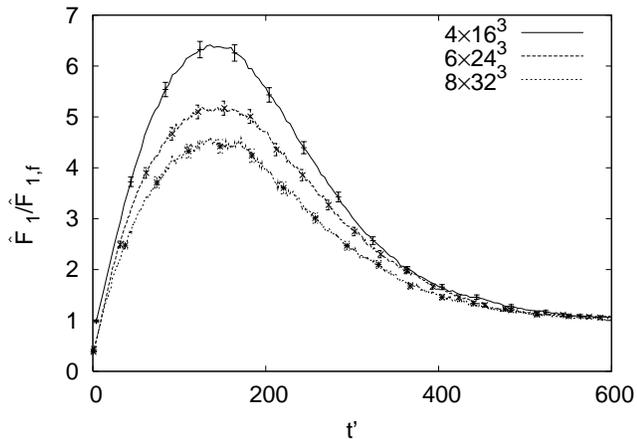,width=\columnwidth} \vspace{-1mm}
\caption{Time evolution of SF $F_1/F_{1,f}$ for SU(3) lattice gauge
theory on $N_{\tau}\,N^3_{\sigma}$ lattices of constant physical
volume of a quench to $T_f/T_c=1.25$. \label{fig_SU3Ntau1} }
\end{center} \vspace{-3mm} \end{figure}

\begin{figure}[-t] \begin{center} 
\epsfig{figure=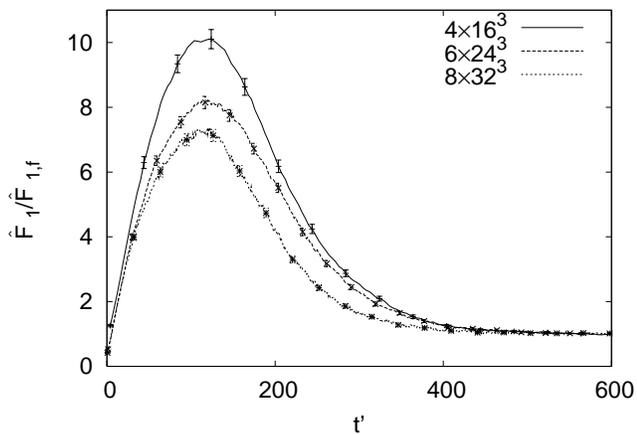,width=\columnwidth} \vspace{-1mm}
\caption{Time evolution of SF $F_1/F_{1,f}$ for SU(3) lattice gauge
theory on $N_{\tau}\,N^3_{\sigma}$ lattices of constant physical
volume of a quench to $T_f/T_c=1.568$. \label{fig_SU3Ntau2} }
\end{center} \vspace{-3mm} \end{figure}

In the following we illustrate the approach of the limit $a\to 0$, 
$L={\rm constant}$, $T_f/T_c = {\rm constant}$, by increasing 
$N_{\tau}$ from 4 to 6 to 8 and the volume $N_{\sigma}^3$ from 
$N_{\sigma}=16$ to 24 to 32, so that the ratio $N_{\sigma}/N_{\tau}$ 
stays constant. The ratio of temperatures $T_f/T_c$ is kept constant
by using Eq.~(\ref{Lambda_L}) to determine the appropriate bare 
coupling constants values for each $N_{\tau}$. Due to the divergence 
of (bare) Polyakov loop correlations we face a renormalization problem, 
which we overcome by dividing all SFs $F_i$ by their equilibrium values 
at $T_f$, $F_{i,f}$. The time-scale situation changes too, because 
we have to use different bare coupling constants values for different
$N_{\tau}$. As one knows that a finite physical volume equilibrates in 
a finite time, we fix this normalization problem by rescaling the time 
axis to
\begin{equation} \label{tp}
  t'\ =\ \frac{t}{\lambda_t(N_{\tau},T_f/T_c)}
\end{equation} 
so that all maxima fall on top of one another. We do not lose 
information as we anyhow do not know the overall normalization 
factor for our time scale. 

Figures~\ref{fig_SU3Ntau1} and~\ref{fig_SU3Ntau2} show the time 
evolution of the $F_1/F_{1,f}$ SFs for our two $T_f/T_c$ values. 
The time axis of our original measurements in units of sweeps are
related to those used in Fig.~\ref{fig_SU3Ntau1} by the 
$\lambda_t(N_{\tau},1.25)$ factors $1:2.655:5.457$ for $N_{\tau}$ 
the values $4:6:8$, respectively. For Fig.~\ref{fig_SU3Ntau2} the 
corresponding $\lambda_t(N_{\tau},1.568)$ ratios are $1:2.768:6.362$.
The maxima of the curves decrease when increasing $N_{\tau}$ from 4
to 6 to 8. As the decrease slows down with increasing lattice size,
there is some evidence for an approach to a shape, which represents
the continuum limit.

\subsection{Debye Screening Mass \label{sec_Debye}}

The current understanding of the early time evolution of systems out
of equilibrium is largely based on investigating stochastic equations
which are dynamical (time dependent) generalizations of the 
Landau-Ginzburg effective action models of the static (equilibrium)
theory~\cite{ChLu97,GD85}. For model~A the linear approximation results
in the following equation for a SF:
\begin{equation}
  \frac{\partial \hat{F}(\vec{p},t)}{\partial t}
  = 2\,\omega(\vec{p})\,\hat{F}(\vec{p},t)\,,
\end{equation}
with the solution
\begin{eqnarray} \label{str_fact}
  \hat{F}(\vec{p},t) \,=\, \hat{F}(\vec{p},t=0)\exp\left(
  2\omega(\vec{p})t\right)\,, \\ \nonumber
  \omega(\vec{p}) > 0 ~~{\rm for}~~|\vec{p}|>p_c\,,
\end{eqnarray}
where $p_c>0$ is a critical momentum. Originally the linear theory was
developed for model~B \cite{CaHi58,Ca68}. Details for model~A can be
found in Ref.~\cite{BMV,Ve04}.

\begin{figure} 
\centerline{\psfig{file=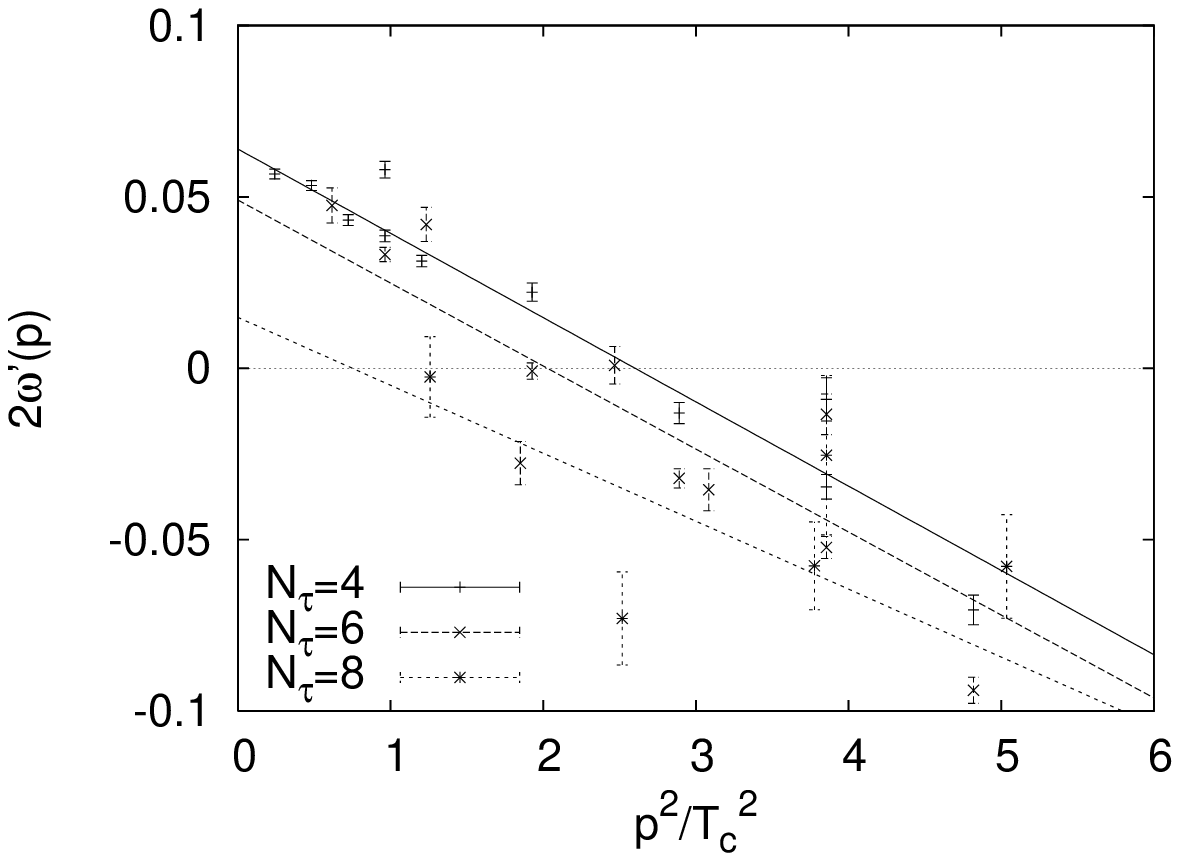,width=9cm}} \vspace*{8pt}
\caption{SU(3) determination of $p_c$  for $T_f/T_c=1.25$.} 
\label{fig_su3kc1} \end{figure}
 
\begin{figure} 
\centerline{\psfig{file=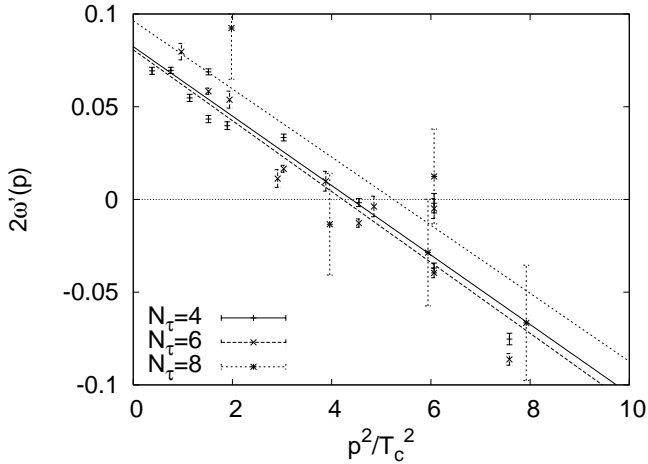,width=9cm}} \vspace*{8pt}
\caption{SU(3) determination of $p_c$  for $T_f/T_c=1.568$.} 
\label{fig_su3kc2} \end{figure}

From our measurements of $F(\vec{p},t)$ on the $N_{\tau}=4$, 
6 and~8 lattices we find straight line 
fits to the form $\omega(p)=a_0+a_1\,p^2,\, p=|\vec{p}|$ with a 
negative slope $a_1$. They determine the critical momentum $p_c$ 
as the value where $\omega(p)$ changes its sign. The fits for 
$T_f/T_c=1.25$ are shown in Fig.~\ref{fig_su3kc1} and for 
$T_f/T_c=1.568$ in Fig.~\ref{fig_su3kc2}, where we introduced
\begin{equation} \label{wp}
  \omega'(p)\ =\ \lambda_t(N_{\tau},T_f/T_c)\,\omega(p)\ .
\end{equation}
This definition absorbs the shift (\ref{tp}) of the time scale, 
so that $\omega'(p)\,t'=\omega(p)\,t$ holds. It is only in the
primed variables that one realizes an eventual approach to the 
continuum limit from Figs.~\ref{fig_su3kc1} and~\ref{fig_su3kc2}. 
In particular note that 
for $T_f/T_c=1.568$ the $N_{\tau}=6$ and~8 fits are within statistical
errors identical. The obtained values for $p_c(N_{\tau})/T_c$ are 
listed in table~\ref{tab_pc}. The (finite volume) continuum limit 
is extrapolated by fitting these values to the form
\begin{equation} \label{pc}
  \frac{p_c(N_{\tau})}{T_c}\ =\ \frac{p_c}{T_c} 
                             +  \frac{\rm const}{N_{\tau}}
\end{equation}
with the results given in the last column of table~\ref{tab_pc}.

\begin{table}[tf]
\caption{Fit results for $p_c/T_c$. \label{tab_pc}} \medskip
\centering
\begin{tabular}{|c|c|c|c|c|}   \hline
Lattice size:&$N_{\tau}=4$   &$N_{\tau}=6$&$N_{\tau}=8$& $\infty$
\\ \hline
$T_f/T_c=1.25:$ & 1.613 (18) & 1.424 (26) & 0.87 (26) & 1.023 (85)\\ \hline
$T_f/T_c=1.568:$& 2.098 (19) & 2.095 (22) & 2.40 (12) & 2.038 (73)\\ \hline
\end{tabular} \end{table} 

Relying on a phenomenological analysis by Miller and 
Ogilvie~\cite{MiOg02}, $p_c$ is related by
\begin{equation} \label{mD}
  m_D\ =\ \sqrt{3}\,p_c 
\end{equation}
to the Debye screening mass at the final temperature $T_f$ after the 
quench. We get
\begin{eqnarray} \label{Debye1}
  m_D &=& 1.77\,(15)\,T_c~~{\rm for}~~T_f/T_c = 1.25\,,\\ \label{Debye2}
  m_D &=& 3.53\,(13)\,T_c~~{\rm for}~~T_f/T_c = 1.568\,.
\end{eqnarray}
The value at $T_f/T_c=1.568$ is in excellent agreement with a
determination of $m_D(T)$ from a best-fit analysis of the large
distance part of the color singlet free energies~\cite{KKZP04}. This 
supports that the simulated dynamics bears physical content. Our 
estimate at $T_f/T_c=1.25$ is by a factor of two smaller than the 
one of Ref.~\cite{KKZP04}. This is not really a surprise, because
$T_f/T_c=1.25$ is close to the spinodal endpoint, so that the
derivation \cite{MiOg02} of the relationship (\ref{mD}) is no 
longer valid.

For pure SU(3) lattice gauge theory $T_c = 265\,(1)\,$MeV holds, 
assuming $\sigma=420\,$MeV for the string tension, while for QCD
the cross-over temperature appears to be around $T_c\approx
165\,$MeV, see Ref.~\cite{Pe05} for reviews. Using for simplicity 
$T_c=200\,$MeV to illustrate the magnitudes, the temporal lattice
size is then about 1~fermi at $T_c$. The spatial sizes of our 
lattices used in this section reach up to $(8\,{\rm fermi})^3$. At 
the $T_f$ values the edge lengths are shortened by the corresponding 
$T_c/T_f$ factors. I.e., the volume is $(6.4\,{\rm fermi})^3$ for 
$T_f/T_c=1.25$ and $(5.10\,{\rm fermi})^3$ for $T_f/T_c=1.568$. The 
screening length associated with the Debye mass, $\xi_D=1/m_D$, is 
then approximately 0.6~fermi at $T_f/T_c=1.25$ and 0.3~fermi at 
$T_f/T_c=1.568$. The illustration of the finite volume continuum
limit in section~\ref{FVCL} was for lattices of size 
$(4\,{\rm fermi})^3$ at $T_c$, i.e., $(3.2\,{\rm fermi})^3$ at 
$T_f/T_c=1.25$ and $(2.55\,{\rm fermi})^3$ at $T_f/T_c=1.568$. 
Our volumes are smaller than the envisioned deconfined region of 
about $(10\,{\rm fermi})^3$ in relativistic heavy ion experiments. 
Due to periodic boundary conditions one may expect that MC simulations 
on smaller lattices are representative for the central region of the 
larger volume. Our result is that the Debye screening length is short 
on the scale of the deconfined region.

\subsection{Measurements near Structure Factor Maxima versus 
            Deconfined Equilibrium}

For SU(3) gauge theory the triality of Polyakov loops with respect to 
the $Z_3$ center of the gauge group takes the place of three distinct 
spin orientation. Although a satisfactory cluster definition does not 
exist for gauge theories, the underlying mechanism of competing 
vacuum domains is expected to be similar as in the spin models. 
To study their influence on Polyakov loop correlations and on the 
gluonic energy $\epsilon$ and pressure $p$ densities, we calculate 
these quantities at times $t\le t_{\max}$ as well as at $t>t_{\max}$.
 
\begin{figure} 
\centerline{\psfig{file=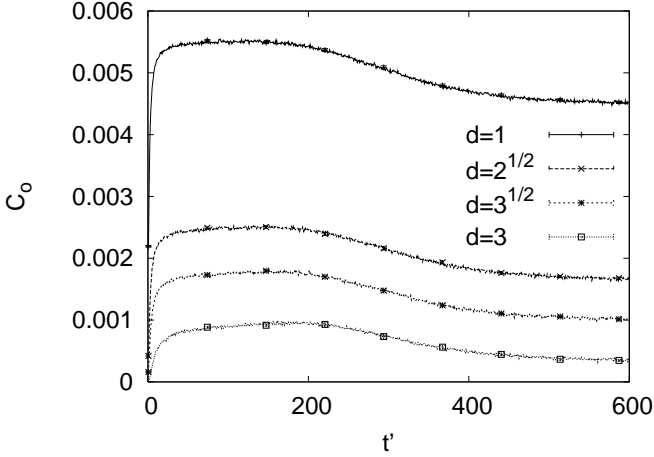,width=9cm}} 
\vspace*{8pt}
\caption{Time dependence of Polyakov loop correlations $C_o(d)$ on the 
      $8\times32^3$ lattice for $T_f/T_c=1.25$ and various $d$ values.} 
\label{fig_su3corrs_all} \end{figure}
 
\begin{figure} 
\centerline{\psfig{file=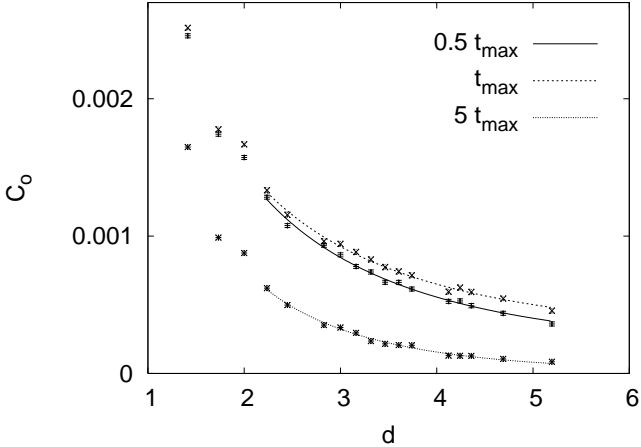,width=9cm}} 
\vspace*{8pt}
\caption{Fall-off behavior of the Polyakov loop correlations of 
         Fig.~\ref{fig_su3corrs_all} at different times. } 
\label{fig_su3corrs_fit} \end{figure}

Our structure function measurements gave ``on the fly'' two-point 
correlations between Polyakov loops defined by
\begin{equation}\label{COR}
   C_o (d,t) = \langle P(0,t)P(d,t) \rangle_L - 
             \left(\langle |P(0,t)| \rangle_L\right)^2
\end{equation}
where the averaging procedures are those we discussed after
Eq.~(\ref{e:L0LR_def}) and $d=1,2,\sqrt{2},\dots$~. The value of 
these results is somewhat limited, because our focus was not on 
good equilibrium results and the stored data do not allow to 
project onto particular channels of the free energy of static 
quarks (which lead to larger correlation lengths than those
obtained). For several 
values of $d$ we plot in Fig.~\ref{fig_su3corrs_all} the time 
development of $C_o(d)$ on our largest lattice using the $t'$ time 
scale (\ref{tp}). The correlations assume maxima at about the same 
time values $t_{\max}$ for which the SFs peak, although less 
pronounced. In Fig.~\ref{fig_su3corrs_fit} we plot the $d$-dependence 
for the time values $0.5\,t_{\max}$, $t_{\max}$ and $5\,t_{\max}$. At 
$5\,t_{\max}$ fits of the form $C_o (d)\sim \exp(-m_P ad)/(ad)$,
where $a$ is the lattice spacing, give the $m_P$ estimates which are 
collected in table~\ref{tab_mP}.  The last column of this table gives 
infinite volume estimates obtained from fits of the form (\ref{pc}).
In contrast to that large correlations are found at $0.5\,t_{\max}$ and 
$t_{\max}$, which are fully consistent with a power law.

\begin{table}[tf]
\caption{Fit results for $m_P/T_c$ at $5\,t_{\max}$. \label{tab_mP}} 
\medskip
\centering
\begin{tabular}{|c|c|c|c|c|}   \hline
Lattice size:&$4\times 16^3$&$6\times 24^3$&$8\times 32^3$& $\infty$
\\ \hline
$T_f/T_c=1.25:$ & 3.27 (19) & 3.70 (20) & 4.43 (27) & 5.23 (49)\\ \hline
$T_f/T_c=1.568:$& 4.82 (61) & 5.33 (35) & 6346 (38) & 7.70 (95)\\ \hline
\end{tabular} \end{table} 

The equilibrium procedure for calculating the gluonic energy $\epsilon$ 
and pressure $p$ densities is summarized in Ref.~\cite{BoEn96,EnKa00} 
(in earlier work \cite{De89} the pressure exhibited a 
non-physical behavior after the deconfining transition and the 
energy density approached the ideal gas limit too quickly because 
the anisotropy coefficients were calculated perturbatively). We denote 
expectation values of space-like plaquettes by $P_\sigma$ and those 
involving one time link by $P_\tau$. The energy density and pressure 
can then be cast into the form
\begin{equation}\label{e:eplusp}
  \frac{\epsilon+p}{T^4} = \frac{8N_cN_\tau^4}{g^2}
  \left[1-\frac{g^2}{2} [c_\sigma(a)-c_\tau(a)]\right] 
  (P_\sigma-P_\tau)
\end{equation}
and
\begin{equation}\label{e:eminus3p}
  \frac{\epsilon-3p}{T^4} = 12N_cN_\tau^4\, 
  \left[ c_\sigma(a)-c_\tau(a) \right]
  \left[2P_0-(P_\sigma+P_\tau)\right] ,
\end{equation}
where $P_0$ is the plaquette expectation value on a symmetric
($T=0$) lattice, and the \textit{anisotropy coefficients}
$c_{\sigma,\tau}(a)$ are defined by:
\begin{equation}\label{e:cst}
    c_{\sigma,\tau}(a)\equiv
    \left(\frac{\partial g^{-2}_{\sigma,\tau}}
    {\partial \xi}\right)_{\xi=1}.
\end{equation}
\begin{figure}
\centerline{\psfig{file=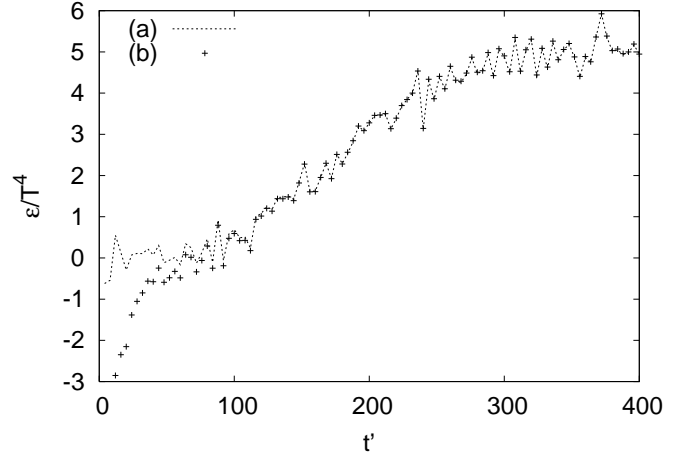,width=9cm}} \vspace*{8pt}
\caption{SU(3) gluonic energy density on the $4\times 16^3$ lattice: 
(a)~with $P_0$ calculated from the time series after the quench and 
(b)~using equilibrium values for $P_0$.} \label{fig_su3subtract}
\end{figure}
They are related to the QCD $\beta$-function and can be calculated
using Pad\'e fits of~\cite{BoEn96}. To normalize to zero temperature, 
plaquette values from the symmetric $N_{\tau}=N_{\sigma}$ lattice are 
needed in Eq.~(\ref{e:eminus3p}). As one stays within the confined 
phase on the symmetric lattice its equilibration after the quench is 
fast.  Therefore it is enough to use equilibrium values of $P_0$ at
$\beta_f$ after the quench. This is illustrated in 
Fig.~\ref{fig_su3subtract}.

\begin{figure}
\centerline{\psfig{file=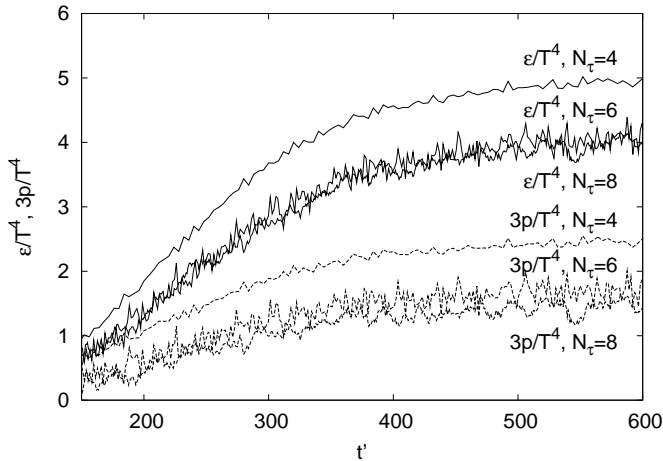,width=9cm}} \vspace*{8pt}
\caption{SU(3) gluonic energy densities and pressures at 
        $T_f/T_c=1.25$.} \label{fig_gedp}
\end{figure}

In Fig.~\ref{fig_gedp} we show the time evolution of the gluonic energy 
densities (upper curves) and pressure densities (lower curves) for 
the $T_f/T_c=1.25$ quench on our $4\times 16^3$, $6\times 24^3$ and 
$8\times 32^3$ lattices using the rescaled time definition~(\ref{tp}). 
The curves for the last two lattices fall almost on top of one another, 
indicating their neighborhood to the continuum limit. The approach to 
the final equilibrium values is rather smooth. Gluonic energy density 
mean values at $t_{\max}$ are less than 1/4 of their final values,
while the pressure density is at about 1/3. In contrast to the 
shift in the mean value, the width of the distributions are almost
the same at $t_{\max}$ and in the deconfined equilibrium. Results
for the $T_f/T_c=1.568$ quench are quite similar.

\section{Summary and Conclusions \label{sec_sum}}

In equilibrium at temperatures much higher than the deconfinement 
temperature $T_c$ the perturbative prescription of QCD is that of 
a weakly coupled gas of quasi-particles. In contrast to that 
recent experiments at the BNL relativistic heavy-ion collider 
(RHIC) show coherence in particle production and strong collective 
phenomena, which are well described by the model of a near-perfect, 
strongly coupled fluid \cite{HK04}. Non-perturbative effects are
expected to play some role in the prescription of equilibrium QCD 
at temperatures reached at the RHIC. For the $T_f/T_c=1.25$ and 
$T_f/T_c=1.568$ temperatures investigated in this paper equilibrium 
lattice calculations indicate indeed corrections (compare Fig.~7 of 
\cite{BoEn96}). However correlations are typically over ranges much 
smaller than the deconfined region, compare our estimates of the 
Debye screening mass $m_D(T_f)$. The agreemet of our $m_D$ value
at $T_f/T_c=1.568$ with direct equilibrium estimates \cite{KKZP04}
give confidence that model~A dynamics reflects physical features.

If the phenomenological description of a strongly coupled plasma
implies correlations over distances exceeding one fermi, the 
time evolution of our structure factors (SFs) depicted in 
Fig.~\ref{fig_3DP3qsf1b30}, \ref{fig_3DP3qsf1b28}, 
\ref{fig_3DP3qtmax} and~\ref{fig_SU3sf1} suggest a scenario 
in which the deconfined equilibrium phase has actually not 
been reached at the RHIC, but the heating process gets stuck 
during the time period of explosive growth of the SFs. 
While this explains correlation over distances much larger than
one fermi, it also provides an {\it unambiguous signal} for the 
existence of the deconfining phase: Fig.~\ref{fig_3DP3qsf1b27} 
demonstrates that the explosive growth is absent for a non-critical 
quench. 

In real QCD there are two effects which prevent the divergence of the 
equilibration time shown in Fig.~\ref{fig_3DP3qtmax}: (1)~Quarks break 
the $Z_3$ symmetry of the SU(3) gauge group, similarly as a magnetic 
field breaks the degeneracy of the spins in the 3D 3-state Potts 
model. The final magnitude of the equilibration time depends then on 
the strength of the breaking as illustrated in Ref.~\cite{BMV} for a 
weak magnetic field. (2)~At the RHIC the physical volume is finite, so 
that even in case of an exact symmetry the equilibration time is 
finite. So the scenario that the system gets stuck during the 
spinodal decomposition of its vacuum structure could 
only be based on phenomenological observations. Questions like how
a perfect fluid may look during the period of spinodal decomposition
arise.  Minkowski space simulations of hyperbolic differential 
equations, which emerge from effective actions for Polyakov loops
\cite{Pi01,Du01}, may shed light on the question whether features
observed in the paper are special to Glauber dynamics or of some
universal nature. 

\acknowledgments
This work was in part supported by the DOE grant DE-FG02-97ER41022 
at FSU and the NSF-PHY-0309362 grant at UCLA. The simulations were 
performed on PC clusters at FSU and UCLA.

\end{document}